\documentclass[twocolumn]{aastex62}
\usepackage{graphicx}
\usepackage{amsmath,amssymb}

\usepackage{graphicx}% Include figure files
\usepackage{dcolumn}% Align table columns on decimal point
\usepackage{bm}% bold math

\usepackage{xcolor}

\newcommand{\CH}{}

\begin{document}

%\preprint{APS/123-QED}

\title{Can Hall Magnetohydrodynamics explain plasma turbulence at sub-ion scales?}

\author[0000-0002-7969-7415]{Emanuele Papini}
 \affiliation{Dipartimento di Fisica e Astronomia, Universit\`a degli Studi di Firenze, via G. Sansone 1, Sesto Fiorentino 50019, Italy}
\email{papini@arcetri.inaf.it}%
\author[0000-0002-7419-0527]{Luca Franci}
\affiliation{School of Physics and Astronomy, Queen Mary University of London, London E1 4NS, United Kingdom}
\affiliation{INAF, Osservatorio Astrofisico di Arcetri, Largo E. Fermi 5, Firenze 50125, Italy}

\author[0000-0002-1322-8712]{Simone Landi}
 \affiliation{Dipartimento di Fisica e Astronomia, Universit\`a degli Studi di Firenze, via G. Sansone 1, Sesto Fiorentino 50019, Italy}
\affiliation{INAF, Osservatorio Astrofisico di Arcetri, Largo E. Fermi 5, Firenze 50125, Italy}

\author[0000-0003-4380-4837]{Andrea Verdini}
 \affiliation{Dipartimento di Fisica e Astronomia, Universit\`a degli Studi di Firenze, via G. Sansone 1, Sesto Fiorentino 50019, Italy}

\author[0000-0002-6276-7771]{Lorenzo Matteini}
\affiliation{LESIA, Observatoire de Paris, Universit\'e PSL, CNRS, Sorbonne Universit\'e, Univ. Paris Diderot, Sorbonne Paris Cit\'e, place Jules Janssen 5, 92195 Meudon, France}

\author[0000-0002-5608-0834]{Petr Hellinger}
\affiliation{Astronomical Institute, CAS, Bocni II/1401, Prague CZ-14100, Czech Republic}

\received{2018 September 6}
\revised{2018 November 5}
\accepted{2018 November 9} 
% \published{2019 January 7}

\begin{abstract}
We investigate the properties of plasma turbulence by means of two-dimensional Hall-magnetohydrodynamic (HMHD) and hybrid particle-in-cell (HPIC) numerical simulations. We find that HMHD simulations exhibit spectral properties that are in most cases in agreement with the results of the HPIC simulations and with solar wind observations. The energy spectra of magnetic fluctuations exhibit a double power-law with spectral index $-5/3$ at MHD scales and $-3$ at kinetic scales, while for velocity fluctuations the spectral index is $-3/2$ at MHD scales. The break between the MHD and the kinetic scales occurs at the same scale in both simulations. In the MHD range the slopes of the total energy and residual energy spectra satisfy a fast Alfv\'en-dynamo balance. The development of a turbulent cascade is concurrently characterized by magnetic reconnection events taking place in thin current sheets that form between large eddies. A statistical analysis reveals that reconnection is qualitatively the same and fast in both the HMHD and HPIC models, characterized by inverse reconnection rates much smaller than the characteristic large-eddy nonlinear time. The agreement extends to other statistical properties, such us the kurtosis of the magnetic field. Moreover, the observation of a direct energy transfer from the large vortices to the small sub-ion scales triggered by magnetic reconnection, further supports the existence of a reconnection-mediated turbulent regime at kinetic scales. We conclude that the Hall-MHD fluid description captures to a large extent the transition of the turbulent cascade between the large MHD scales and the sub-ion scales.

%\newline \vspace*{.5 pt}\newline
%Subject Areas: Astrophysics, Computational Physics, Plasma physics
\end{abstract}

%\pacs{Valid PACS appear here}
%\keywords{Astrophysics, Plasma physics, Magnetic reconnection}%Use showkeys class option if keyword
%Use showkeys class option if keyword
                              %display desired
%\maketitle

\section{INTRODUCTION}

Understanding the mechanisms responsible for transferring magnetic and kinetic energy from large to small scales in weakly collisional plasmas is of great interest in space, astrophysical, and laboratory plasmas, since such mechanisms are relevant in many fundamental processes: the formation of hot coronae, the heating and acceleration of stellar winds and their interaction with planetary magnetospheres, the acceleration of particles, and explosive phenomena such as solar flares and coronal mass ejections, to name a few.
Indeed, in hot rarefied plasmas collisions between particles are not efficients in dissipating energy at scales above the particle's characteristic kinetic scale, namely the ion (electron) Larmor and  demagnetisation scales. 
This is the case of, e.g., the solar wind and the Earth's magnetosheath.
In-situ spacecraft observations with high spatial and temporal resolution \cite[see, e.g.,][]{2013chen_b,2017chen} have shown that the magnetic energy spectrum follows a Kolmogorov's power-law cascade at scales larger than the ion kinetic length scales.
As the cascade approaches these scales, namely the ion inertial length $d_i$ or the ion gyroradius, the one-fluid magnetohydrodynamic description breaks and nonlinear interactions are strongly modified by the different propagation and polarization properties introduced by the specific species's dynamics. For instance,
the heavy ions have a different gyromotion around the magnetic field lines than the light electrons, and at scales below $d_i$ the motion of the electrons decouples from the ion's (Hall effect).

Indeed, observations have shown that around such scales the magnetic field spectrum steepens and the nature of the fluctuations changes \CH{\citep[see, e.g.,][]{1983denskat,1994goldstein}}. 
In particular, a power-law spectrum with a slope between $ -2.8$ and $-3$ is routinely observed \cite[see, e.g.,][]{Alexandrova_al_2009,2017chen}. Concurrently, the electric to magnetic field power increases \citep{Salem_al_2012}, together with compressive fluctuations \cite[see, e.g.,][]{Chen_al_2012,2013chen,2017matteini}, and the nature of the fluctuations seems to be compatible with strongly obliquely Alfv\'enic modes, the so-called kinetic Alfv\'en waves \cite[see, e.g.,][]{Salem_al_2012,2013chen}. 

{Early attempts to describe the transition of the electromagnetic turbulence across ion scales use fluid extended magnetohydrodynamic (MHD) models, either in the low (Hall-MHD) \CH{\citep{1996ghosh,2006galtier,2007galtier,2009shaikh} or in the high (Electron MHD) \citep{Biskamp_al_1996,Biskamp_al_1999,Cho_Lazarian_2004,Cho_Lazarian_2009,2013meyrand}} frequencies approximation. 
In these approaches, we observe a transition near the ion scales with a steepening in the magnetic field power and a simultaneous flattening in the electric field power fluctuations \citep{Matthaues_al_2008}.
Steeper spectra, as observed in the solar wind and in magnetospheres, have been often reproduced by the use of both 2D \citep{2015franci_a,2015franci_b,2016franci} and 3D \cite[see, e.g.,][]{2011howes,Franci_al_2018} kinetics models, as well as in fluid models that include kinetic dissipative effects like the ion and electron Landau damping \citep{Sulem_al_2016}.  
It has been thus suggested than the steepening of the spectra is strictly correlated to deformations in the particle's velocity distribution function associated, for example, to wave-particle interactions such as instabilities and wave damping, to thermal anisotropies,  and/or to non gyrotropic terms in the full pressure tensor \citep{DelSarto_al_2016b,DelSarto_Pegoraro_2018} that can drive dissipative effects \citep{Yang_al_2017a}.
Recently, high resolution hybrid particle-in-cell (HPIC) simulations \citep{2015franci_a,2015franci_b}, that use a kinetic description for the ions and a fluid one for the electrons, were able to reproduce the steep spectra observed at kinetic scales in the solar wind and to show that the transition between the MHD and the ion scales is correlated to the largest characteristic kinetic scale encountered by the nonlinear cascade \citep{2016franci}. It was also shown that the formation of the turbulent cascade in the sub-ion range is associated to the formation of small-scale reconnecting current sheets, \CH{which rapidly transfer the magnetofluid energy at sub-ion scales \citep{2017franci,2017cerri}}. Finally, analyses of HPIC simulations based on the von Karman-Howarth equations \CH{\citep{2008galtier,2018hellinger} indicate that a part of the MHD cascade
continues at sub-ion scales via the Hall term.
The two cascades are not, however, separable and overlap
(the Hall part progressively dominating at smaller scales),
both contributing to the total Hall-MHD cascade.}

The main goal of this work is to check whether a fluid model that allows for the formation of small scales coherent structures, such as current sheets, can reproduce spectra steeper than what expected by simpler phenomenological models, without the need to include purely kinetic effects. 
To that purpose, we performed a 2D high resolution Hall-MHD (HMHD) simulation for a warm polytropic plasma and compared the outcome with analogous results obtained from a HPIC simulation. 

Our results suggest that the main processes for the formation of steeper spectra  involve the formation of coherent small-scales structures and are consistent with \CH{recent models of reconnection-mediated turbulence},  such as that proposed by \cite{Boldyrev_Perez_2012}, \cite{2017loureiro}, and \cite{2017mallet}.

\section{Hall-MHD simulations: numerical setup}
Our model integrates the viscous-resistive MHD equations but retaining the Hall term in the induction equation, that is, by substituting the fluid velocity $\boldsymbol{v}$ with the electron velocity $\boldsymbol{v}_e = \boldsymbol{v} - \boldsymbol{J}/ e n_e$. In their adimensionalized form, the HMHD equations read
\begin{align}
 \partial_t{\rho} + \boldsymbol{\nabla}\cdot{(\rho \boldsymbol{v})} = & ~0,
  \label{eq:continuity}
  \\
 \rho\left (\partial_{t} + \boldsymbol{v}\cdot\boldsymbol{\nabla}\right) \boldsymbol{v} = & -\boldsymbol{\nabla} P + (\boldsymbol{\nabla}\times{\boldsymbol{B}}) \times\boldsymbol{B} \nonumber \\
  & + \nu \left [ \Delta\boldsymbol{v} + \frac{1}{3}\boldsymbol{\nabla}(\boldsymbol{\nabla}\cdot{\boldsymbol{v}}) \right],\\
  \left (\partial_{t} + \boldsymbol{v}\cdot\boldsymbol{\nabla}\right) T = &
  ~(\Gamma  -  1) \! \left \{ - (\boldsymbol{\nabla}\cdot{\boldsymbol{v}})T 
  \! + \!  \eta \frac{|\boldsymbol{\nabla}\times\boldsymbol{B}|^2}{\rho} \right . \nonumber\\
  & + \left . \frac{\nu}{\rho} \left [ (\boldsymbol{\nabla}\times{\boldsymbol{v}})^2  +\frac{4}{3} (\boldsymbol{\nabla}\cdot{\boldsymbol{v}})^2\right ] \right \}
  ,   \\
   \partial_t{\boldsymbol{B}} = & ~\nabla\!\times\!\left ( \boldsymbol{v} \times \boldsymbol{B}\right ) \! + \! \eta \nabla^2 \boldsymbol{B} \nonumber\\
 & -  \eta_H\nabla\!\times\! \frac{(\nabla\!\times\!\boldsymbol{B})\!\times\!\boldsymbol{B}}{\rho},
  \label{eq:induction_hall_adi}
\end{align}
where $\Gamma=5/3$ is the adiabatic index and the variables retain their usual meaning. All quantities are renormalized using $L=d_i$ as characteristic length and with respect to a field amplitude $B_0$, a density $\rho_0$, an Alfv\'en speed $c_A = B_0/\sqrt{4\pi\rho_0} = \Omega_i d_i$, a pressure $P_0=\rho_0 c_A^2$, and a temperature $T_0 = (k_B/m_i) P_0/\rho_0$. Here, the adimensional magnetic resistivity $\eta$ is in units of $d_i c_A$ and the Hall coefficient $\eta_H=d_i/L$ is equal to 1. $\Omega_i = e B_0 / m_i c$ is the ion-cyclotron angular frequency and $m_i$ is the mass of the ions.

The equations (\ref{eq:continuity}-\ref{eq:induction_hall_adi}) are numerically solved by using a code we employed for studies of magnetic reconnection \citep{2015landi,2018papini}, modified to include periodic boundaries in all directions.
We consider a 2D $(x,y)$ periodic domain and use Fourier decomposition to calculate the spatial derivatives. In Fourier space we also filter according to the $2/3$ Orszag rule \CH{\citep{1971orszag}, to avoid aliasing of the nonlinear quadratic terms. Aliasing of the cubic terms is mitigated by the presence of a finite dissipation \citep{1993ghosh}. } Time integration is performed via a 3rd-order Runge-Kutta scheme.
The other code employed in this work is the Lagrangian hybrid particle-in-cell (HPIC) code CAMELIA \citep{1994matthews,2017franci_astronum}\CH{, which integrates the Vlasov-Maxwell equations by coupling fully kinetic ions to a charge-neutralizing and massless fluid of isothermal electrons. CAMELIA} has been successfully used for numerical studies of plasma turbulence \citep{2015franci_a,2015franci_b,2016franci,2016bFranci,2017franci}, and it reproduced many of the spectral properties observed in the solar wind \citep[][]{2017franci_astronum} \CH{and in the Earth's magnetosheath by MMS (Franci et al., in preparation)}.

The numerical setup used in this work is the same as in the HPIC simulation of \cite{2017franci}, in order to allow a straightforward comparison between HMHD and HPIC results.  
We consider a 2D box of size $L_x \times L_y = 256~d_i \times 256~d_i$ and a grid resolution of $\Delta x = \Delta y = d_i/8$, corresponding to $2048^2$ points.
Out of the plane, along the $z$ direction (identified as the parallel direction), we set a background constant magnetic field $B_{0}$, which we refer to as the mean field.
The initial state is populated by freely-decaying large-amplitude \CH{Alfv\'enic-like fluctuations with zero mean cross helicity in the $xy$-plane perpendicular to the mean field \citep{2015franci_b}} and up to the  injection scale $\ell_\text{inj}=2\pi d_i/k_\perp^\text{inj}$, with $k_\perp^\text{inj} d_i \simeq 0.2$, where  $k_\perp = \sqrt{k_x^2+k_y^2}$. 
The root-mean-square (rms) amplitude of these fluctuations is set to $b_\mathrm{rms} = B_\mathrm{rms}/B_0 \simeq 0.24$.
In the HPIC reference simulation, the plasma beta for ions and electrons and the magnetic resistivity were set to the values $\beta_i=\beta_e=1$ and $\eta=5\times 10^{-4}$ respectively. In HMHD we set the plasma $\beta=\beta_i+\beta_e=2$ accordingly. Instead, for the resistivity, we employ $\eta= 10^{-3}$, i.e., twice the value of the hybrid simulation. We also set the viscosity to the same value. \CH{At the global scales of the box, these values correspond to a viscous and magnetic Reynolds number of $R=R_m\simeq 40~000$.}

\section{Role of Magnetic reconnection in developing turbulence}
\begin{figure*}
 \includegraphics[width=\textwidth]{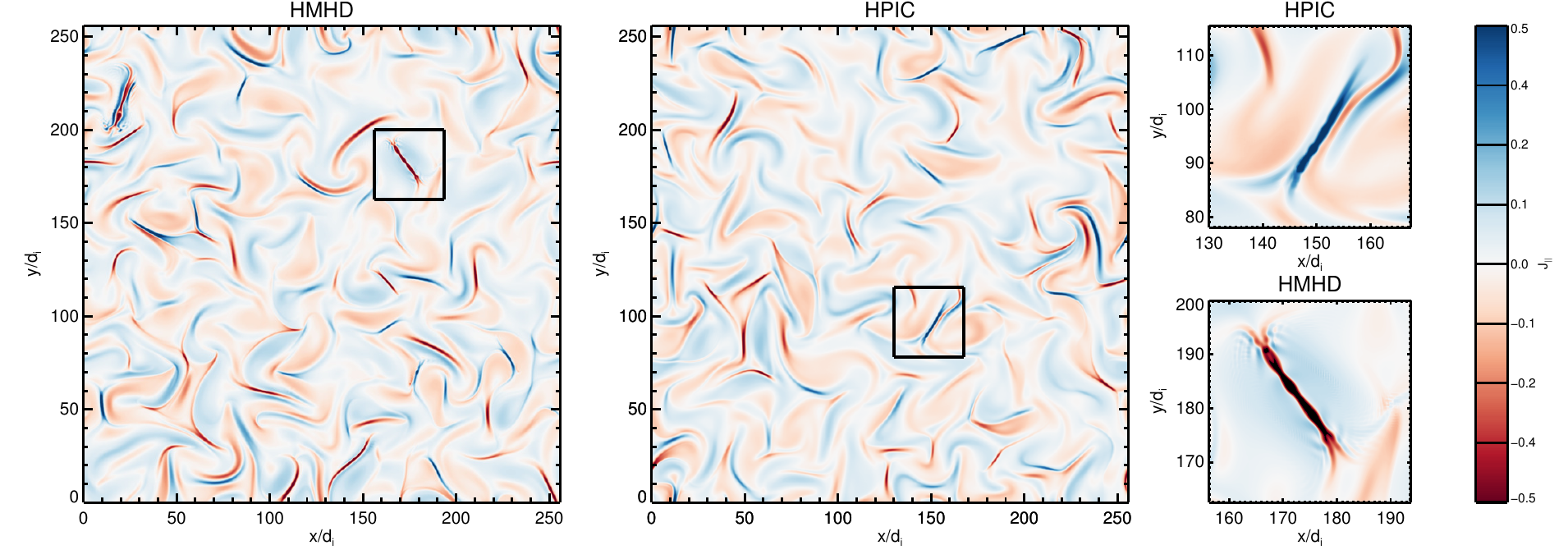}
 \caption{Coloured contours of the out-of-plane current density $\boldsymbol{J}_\parallel$, for the HMHD and the HPIC simulations at \CH{$t= 45~\Omega_i^{-1}$}, in the whole grid (big panels on the left) and in a subgrid containing one reconnecting current sheet (small panels on the right). Color scales are saturated to $\pm 0.5$ for easy visualization of the structures.}
 \label{fig:current_dev}
\end{figure*}
\begin{figure}
    %\centering
    \includegraphics[width=\columnwidth,trim={0.cm 0 1.2cm 0},clip]{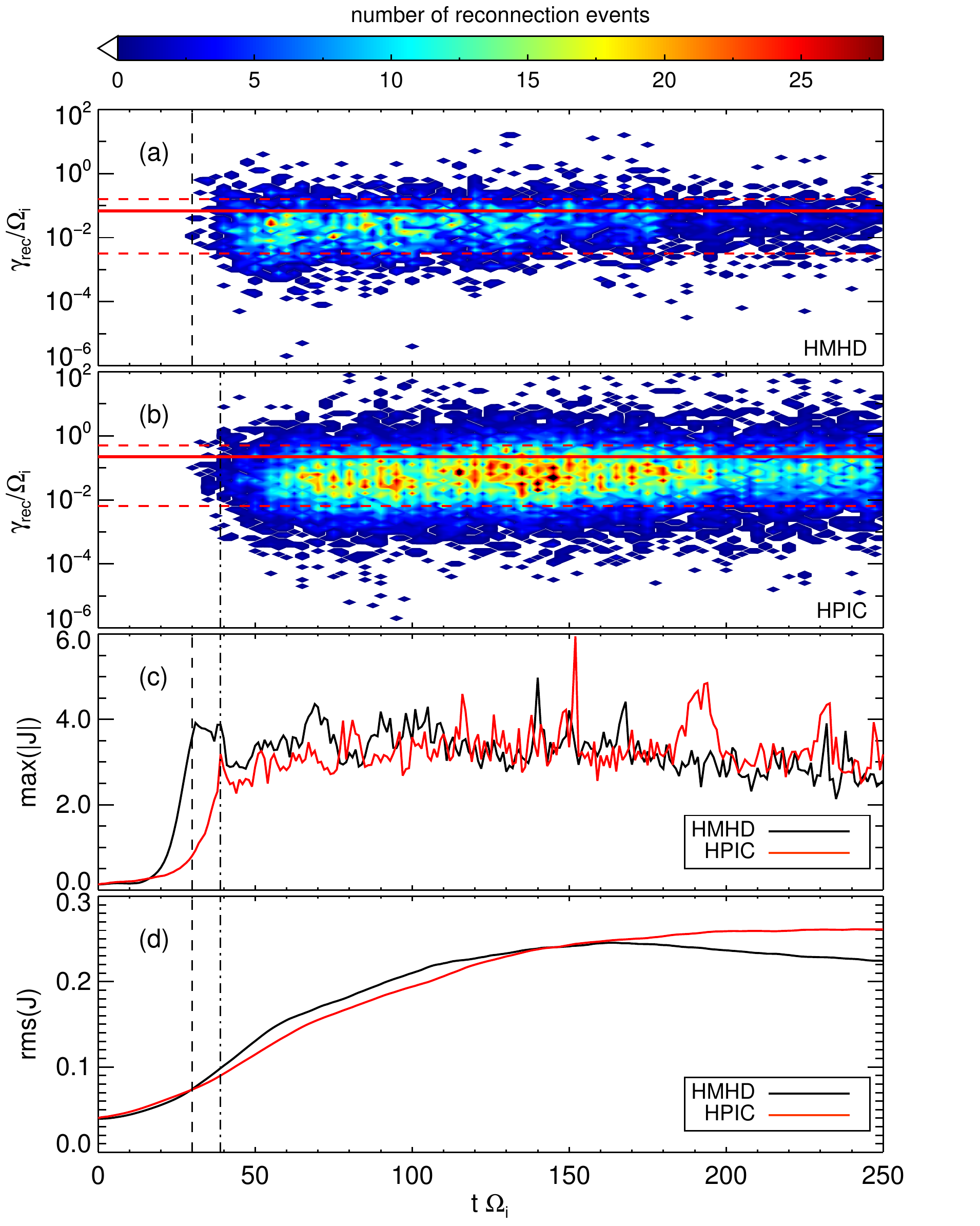}
    \caption{Time evolution of the reconnection rates $\gamma_\text{rec}/\Omega_i$ for the HMHD (a) and the HPIC run (b): coloured contours indicate the distribution of reconnection events, horizontal solid red lines  denote the average reconnection rate, and  the dashed red lines denote $10$th and $90$th percentile values respectively. Time evolution of the maximum of $|\boldsymbol{J}|$ (c), and of the rms value of $\boldsymbol{J}$ (d). In all plots, the vertical dashed and dot dashed lines mark the first maximum in $|\boldsymbol{J}|$ of the HMHD run and the HPIC run respectively.}
    \label{fig:recrates}
\end{figure}

\begin{figure}
   \includegraphics[width=\columnwidth]{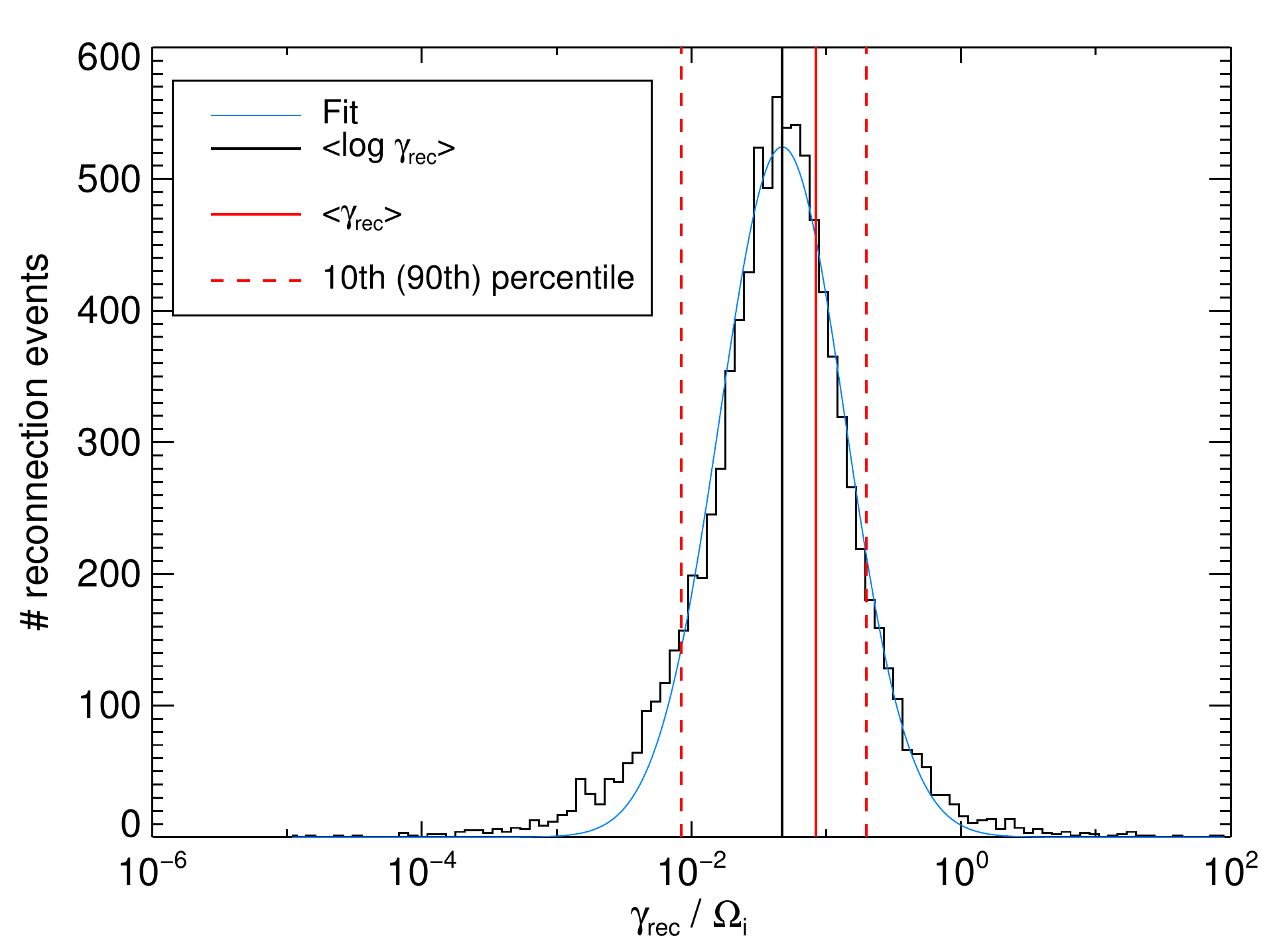}
   \caption{\CH{Histogram of the reconnection rates $\gamma_\text{rec}$ as measured in the HMHD run at all times. 
   The blue curve is the lognormal fit to the histogram, corresponding to a gaussian fit of $\log (\gamma_\text{rec})$.
   As in Fig. \ref{fig:recrates}a, the red solid line denotes the mean reconnection rate, and the dashed red lines denote $10$th and $90$th percentile values respectively. The black solid line denotes the mean of $\log (\gamma_\text{rec})$.}}
    \label{fig:recrates_hist}
\end{figure}

The initial Alfv\'enic fluctuations quickly evolve to form vortices and localized current sheets, which then shrink down to a critical width of the order of $d_i$ \CH{in a time of about $30~\Omega_i^{-1}$.} 
These current sheets quickly disrupt, due to the onset of fast magnetic reconnection processes, generating a variety of small scale magnetic islands and fluctuations that are fed back into the turbulent surrounding \CH{\citep[a similar dynamics is known to occur also in MHD turbulence, see, e.g.,][]{1986matthaeus, 1989biskamp, 1990carbone}}. As an example, Figure \ref{fig:current_dev}  shows a contour plot of the out-of-plane current density, $\boldsymbol{J}_z$, at $t=45~\Omega_i^{-1}$, right after the first reconnection events have been observed, for both the HPIC and the HMHD runs. In particular, a zoom on two current sheets (small panels) reveal a plasmoid-chain structure, a characteristic footprint of the tearing instability \cite[see, e.g.,][]{2007loureiro,2018papini}. 
As the system evolves, other reconnection events take place in newly formed current sheets, until turbulence is fully developed.

To quantitatively support the above description, we analyzed in detail 
the reconnection events in both simulations. A reconnection event is identified by a magnetic X-point and its nearest O-point in a current sheet. For each event we calculated the reconnection rate
\begin{equation}
    \label{eq:recrate}
     \gamma_\mathrm{rec} = \left | \frac{1}{\Phi|_X^O} \frac{\partial \Phi|_X^O}{\partial t} \right |,
\end{equation}
\CH{where $ \Phi|_X^O$ is the reconnected magnetic flux density  between the X-point and the O-point, i.e., the difference in the out-of-plane vector potential between those points, $ \Phi|_X^O = A_z^O - A_z^X$ (for further details, see the Appendix)}.
Figure \ref{fig:recrates} shows the temporal evolution of the distribution of reconnection rates for the HMHD run (a) and the HPIC run (b), together with the maximum of $|\boldsymbol{J}|$ (c) and the root-mean-square (rms) of the current density $\boldsymbol{J}$ (d).
For the HMHD run, the first reconnection events are detected at $t\simeq 30~\Omega_i^{-1}$ and concurrently with the first maximum of $|\boldsymbol{J}|$ (marked by a vertical dashed line), which is a proxy for reconnection events \citep{2017franci}. 
The reconnection rates follow a lognormal distribution, with a mean value of $\langle \gamma_\mathrm{rec}\rangle=0.08~\Omega_i$ (red solid line). \CH{Figure \ref{fig:recrates_hist} shows the distribution of all the reconnection rates measured in the HMHD run at all times.} As the system evolves, this value remains roughly constant and correspond to an average reconnection time $\langle\tau_\mathrm{rec}\rangle =1/\langle \gamma_\mathrm{rec}\rangle \simeq 12.2~\Omega_i^{-1}$. 
We note however that a significant fraction (roughly $22\%$) of the reconnection events has a reconnection time smaller than the average, and $10\%$ of the reconnection events are very fast, with $\tau_\mathrm{rec}  < 6~\Omega_i^{-1}$. 
After the maximum of turbulent activity is reached, i.e., after the maximum of rms($|\boldsymbol{J}|$) \citep{2009mininni} at $t\simeq 165~\Omega_i^{-1}$, the number of reconnection events decreases.
\begin{figure}
    %\centering
    \includegraphics[width=\columnwidth,trim={0.cm 0 0.5cm 0},clip]{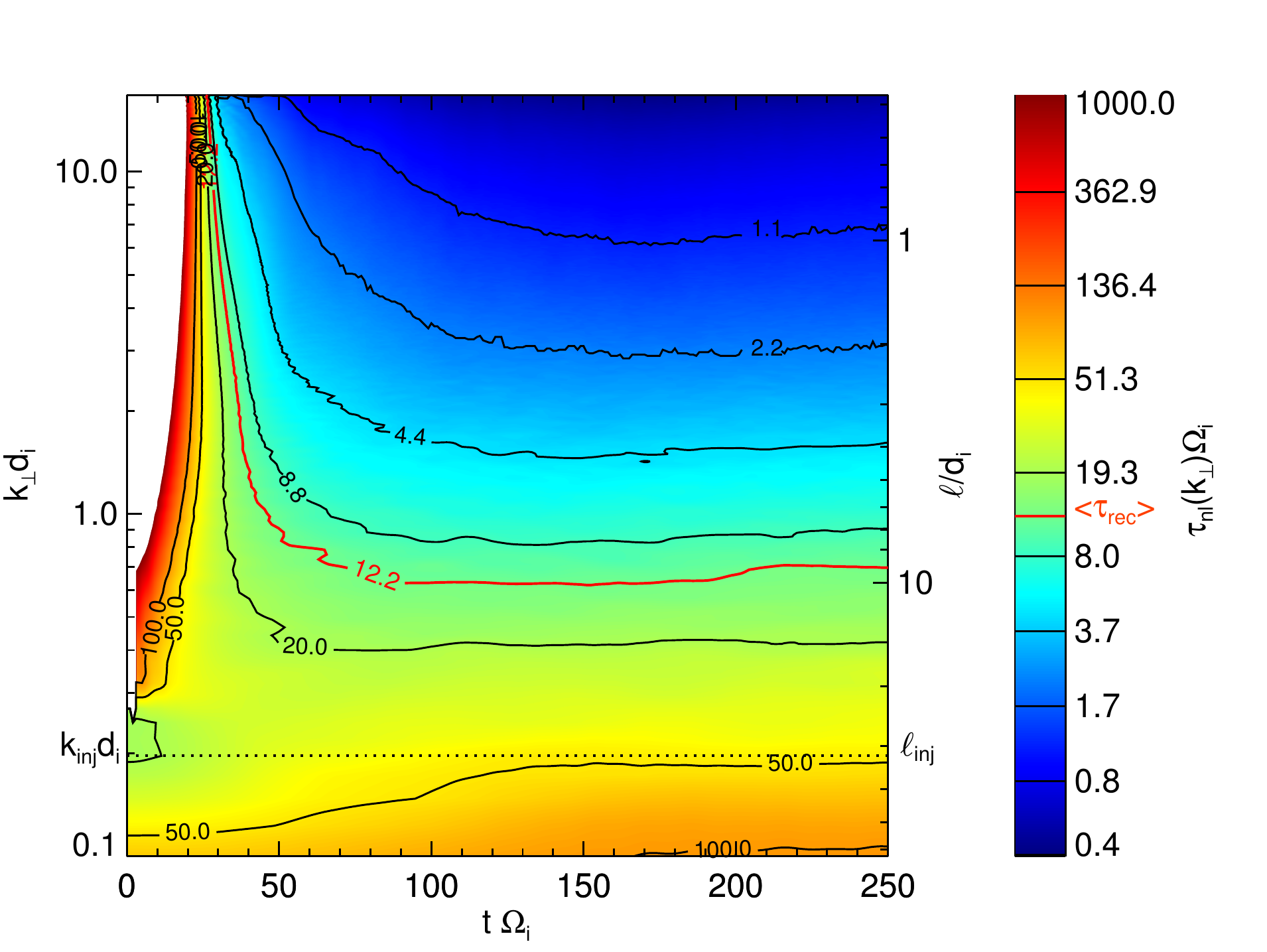}%[width=\columnwidth]
    \caption{Coloured contours of the characteristic nonlinear time $\tau_{nl}(k_\perp)$ as a function of wavenumber $k_\perp$ (scale $\ell$) and time $t$.
    The right axis denotes the corresponding scale $\ell=2\pi/k_\perp$.
    The red contour line highlights the scale at which the nonlinear time equals the average reconnection time. The white area denotes the region where the nonlinear time exceeds $1000~\Omega_i^{-1}$. The horizontal dotted line indicates the injection scale.}
    \label{fig:nltimes_vs_recrates}
\end{figure}

Magnetic reconnection modifies the turbulent properties at kinetic scales soon after the first reconnection events are detected. This is made evident by looking at the nonlinear time associated to turbulence, that is the eddy turnover time. Such a characteristic time at a given scale is estimated as 
$\tau_{nl}(k_\perp) = (k_\perp v_e (k_\perp))^{-1} $, where $v_e(k_\perp)$ is the electron velocity at the scale $\ell=2\pi/k_\perp$. The use of this definition unifies the classic definition at MHD scales (where the electron velocity equals the fluid velocity) and the definition of the nonlinear time at kinetic scales.
Figure \ref{fig:nltimes_vs_recrates} shows a contour plot of the nonlinear time as a function of time and wavenumber. 
Here one can identify three distinct phases: an early evolutionary phase ($0\leq t\Omega_i\lesssim30$) in which the initial configuration relaxes and the first vortices and current sheets are formed, a transient phase ($30\lesssim t\Omega_i\lesssim50$) triggered by the first reconnection events, and a third phase ($50\lesssim t\Omega_i$) characterized by slowly evolving values of $\tau_{nl}(k_\perp)$ that lasts until the end of the simulation. 
In the first phase, at the beginning, the scale with the smallest value of $\tau_{nl}({k_\perp})$ is the injection scale $\ell_\mathrm{inj}$ (marked by the horizontal dotted line)\CH{, with $\tau_{nl}({k_\mathrm{inj}}) \simeq 22~\Omega_i^{-1}$.} As the system evolves and the energy gets redistributed, the nonlinear time at scales above the injection scale increases, while at the scales right below the injection scale the nonlinear time slightly decreases, due to the development of a direct cascade. At small scales and  before the first reconnection events are detected ($ t \lesssim 30~\Omega_i^{-1}$), the nonlinear time is very large, since there is no relevant amount of energy at those scales yet.
As soon as reconnection is triggered, the second phase begins. The energy is directly transfered to the smallest scales accessible (the 2/3 cutoff scale), where $\tau_{nl} (k_\perp)$ suddenly decreases. Then, during a transient phase that last between $t \simeq 30~\Omega_i^{-1}$ and $t \simeq 50~\Omega_i^{-1}$ (characterized by almost vertical isocontour lines of $\tau_{nl}(k_\perp)$. See, e.g., the red contour line), the energy is fed to larger and larger scales. We interpret \CH{this behavior as the signature in Fourier space of the
coalescence of magnetic islands} \citep{1977finn}, which we also observe in the real space at the reconnection sites. During this transient, the number of reconnection events increases very rapidly and reaches a statistically stationary value at about $t=50~\Omega_i^{-1}$. After this transient, the nonlinear time $\tau_{nl}(k_\perp)$ changes only slightly and then becomes roughly constant at $t\gtrsim 150~\Omega_i^{-1}$, indicating that a stationary turbulent state has been reached (see Section \ref{sec:spectra}).
Figure \ref{fig:nltimes_vs_recrates} provides a direct quantitative evidence of the ability of magnetic reconnection to influence the dynamics of turbulence at kinetic scales. 

The evolution in the HPIC simulation is qualitatively the same, though the first reconnection events are triggered at a slightly later time ($t\simeq 40~\Omega_i^{-1}$) and the distribution of the reconnection rates is broader, with an average reconnection rate  
$\langle \gamma_\mathrm{rec}\rangle=0.22~\Omega_i$, roughly three times the one of the HMHD run. 
We believe that nongyrotropic ion effects decrease the amplitude of the out-of-plane ion velocity $\boldsymbol{v}_{i,\parallel}$ at the reconnection sites in the HPIC simulation, thus increasing the reconnection rate, as demonstrated by \cite{2002yin}. Indeed, in the current sheets of Fig.\ref{fig:current_dev} we measured an amplitude of $\boldsymbol{v}_{i,\parallel}$ at the X-points  in the HPIC run that is roughly 1/3 of that in the HMHD run.

Note that the maximum of turbulent activity in the HPIC run is reached later, at about $t\simeq 200~\Omega_i^{-1}$, due to the smaller value of the resistivity employed, which sets a dissipation scale smaller than the one in the HMHD run, thus increasing the time needed to fully develop a turbulent cascade.

\section{Intermittency and spectral properties in fully developed turbulence}
\label{sec:spectra}

From the previous section we conclude that, despite the differences in the numerical approaches and the theoretical models, the HMHD and the HPIC simulations have remarkable similarities. The most interesting agreement is found in the  spectral properties.

Figure \ref{fig:spectra_b} shows the omnidirectional power spectra of magnetic and velocity fluctuations of both the HPIC and the HMHD run, at the time when turbulence has fully developed and the rms of $\boldsymbol{J}$ reached its maximum. We remind that the fluid velocity $\boldsymbol{v}$ in the HMHD description corresponds to the ion bulk velocity $\boldsymbol{v}_i$ of the HPIC model. After the maximum, the spectra remain quite stable.
\begin{figure}
    %\centering
    \includegraphics[width=.95\columnwidth,trim={0.cm 0 0.4cm 0},clip]{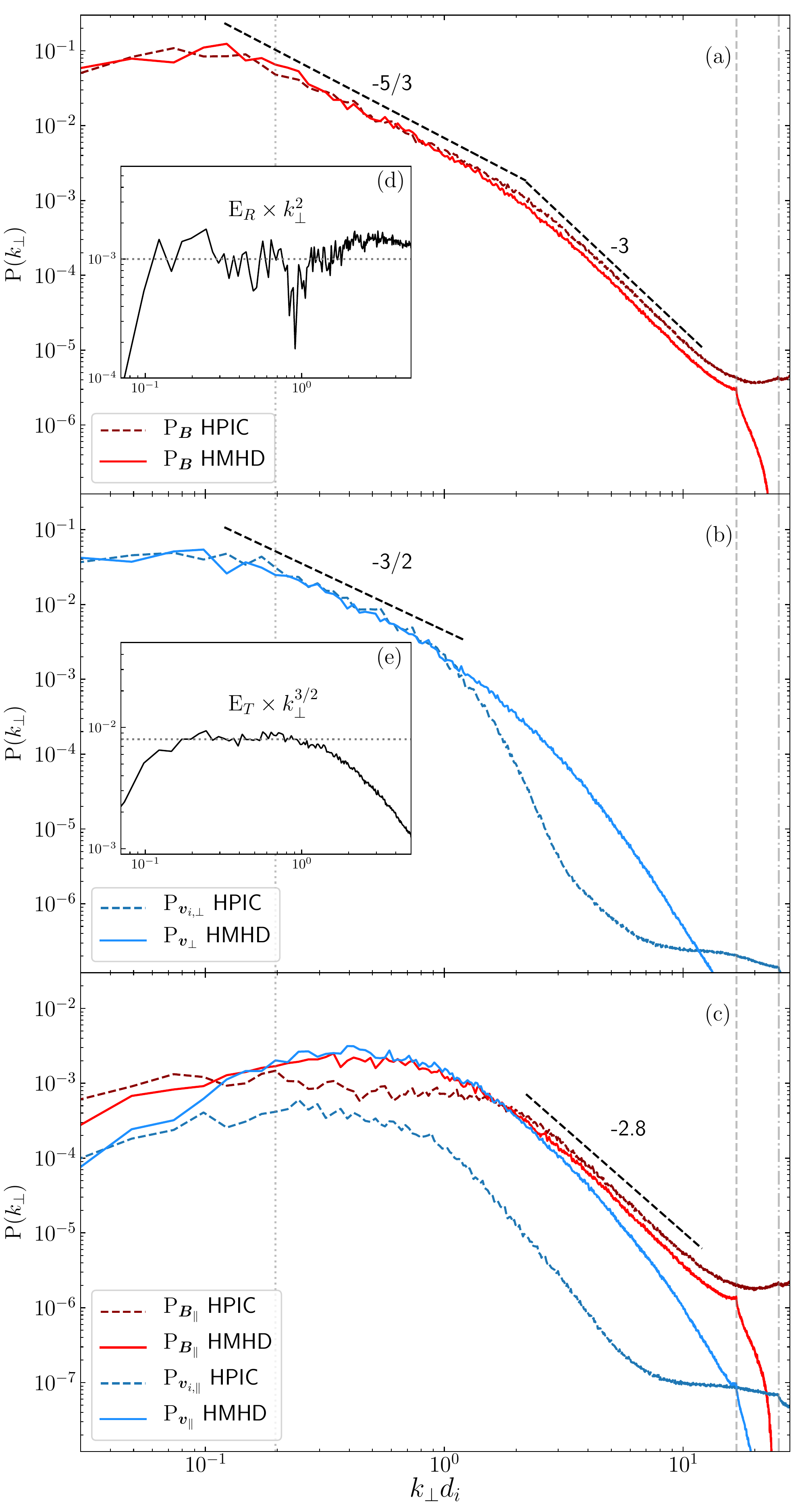}
    \caption{Isotropized power spectra $\mathrm{P}(k_\perp)$ as a function of $k_\perp = \sqrt{k_x^2+k_y^2}$ from the HMHD run (solid curves) at $t=165~\Omega_i^{-1}$ and from the HPIC run (dashed curves) at $t=200~\Omega_i^{-1}$, of total magnetic field (a) and perpendicular ion velocity (b) fluctuations, and of parallel magnetic field and parallel velocity fluctuations (c). The vertical dotted, dashed, and dot-dashed lines denote the injection scale $k_{\perp}^\mathrm{inj} d_i \simeq 0.2$, the 2/3 filter's cutoff of the HMHD model, and the Nyquist wavenumber respectively. Black dashed lines denote Reference slopes. Panel (d) shows the residual energy spectrum $\mathrm{E}_R = \mathrm{P}_{\boldsymbol{B}} - \mathrm{P}_{\boldsymbol{v}} $, compensated by $k_\perp^{2}$, from the HMHD simulation.}
    \label{fig:spectra_b}
\end{figure}
At MHD scales, we observe a Kolmogorov-like cascade with slope $-5/3$ in the magnetic field fluctuations (see Fig. \ref{fig:spectra_b}(a)), while the power spectrum of the fluid (ion) velocity field fluctuations is flatter, with a spectral index of about $-3/2$ (see  Fig. \ref{fig:spectra_b}(b)). This is in agreement with solar wind observations \citep{2011chen}.

A transition occurs at ion kinetic scales, where the magnetic field steepens following a power-law of $-3$.
The HPIC and HMHD power spectra of total magnetic field fluctuations (Fig. \ref{fig:spectra_b}(a)) almost perfectly match at all scales, from the injection scale (vertical dotted line) down to the scale where the HMHD spectrum has the filter's cutoff at 2/3 of the Nyquist frequency (vertical dashed line) and the HPIC spectrum rises due to numerical noise.  
The location of the spectral break at $k_\perp d_i \simeq 2$ also matches.
The agreement extends to the spectra of the perpendicular magnetic fluctuations (not shown here).

For the spectra of parallel magnetic fluctuations (Fig. \ref{fig:spectra_b}(c)) we find some compelling differences.
At kinetic scales, the parallel magnetic field spectra have the same quantitative and qualitative behavior, though the HPIC spectrum of $\boldsymbol{B}_{\parallel}$ is compatible with a power law of spectral index $-2.8$, while the HMHD spectrum of $\boldsymbol{B}_{\parallel}$ is more compatible with a value of $-3$. 
At MHD scales the two spectra are different. There the HMHD spectrum of $\boldsymbol{B}_{\parallel}$ is coupled to the spectrum of parallel velocity fluctuations (blue solid curve), thus suggesting that parallel Alfv\'enic fluctuations give the dominant contribution. In the HPIC case (dashed curves) such a coupling is not present. 
This picture is reversed in the spectra of the perpendicular velocity fluctuations (Fig. \ref{fig:spectra_b}(b)):
at MHD scales and for both the HMHD and the HPIC simulations we recover a power law cascade of spectral index $-3/2$.
At kinetic scales ($k_\perp d_i >1$) the two spectra diverge, due to the lacking ability of Hall-MHD to model ion kinetic effects.

In the MHD range of the HMHD run ($0.3\lesssim k_\perp d_i\lesssim1$) 
we obtain a residual energy spectrum $\mathrm{E}_R = \mathrm{P}_{\boldsymbol{B}} - \mathrm{P}_{\boldsymbol{v}}$ consistent with a slope of -2 (Fig. 4(d)) and a total energy spectrum $\mathrm{E}_T = \mathrm{P}_{\boldsymbol{B}} + \mathrm{P}_{\boldsymbol{v}} \propto k_\bot^{-3/2}$ (Fig. 4(e)). While the former agrees with the HPIC results \citep{2015franci_a}, the higher energy in the parallel component of velocity fluctuations found in the HMHD run causes its total energy spectrum to be flatter than $-5/3$.
\cite{2005mueller} and
\cite{2016grappin} proposed that the residual energy spectrum and the total energy spectrum are related by a balance between a local dynamo effect and an Alfv\'enic coupling. Their relation reads: 
\begin{equation}
{\mathrm{E}_R}/{\mathrm{E}_T}\approx A \left({t_A}/{t_{nl}}\right)^\alpha
\label{eq:dyn}
\end{equation}
where the Alfv\'en time $t_A=1/(k_\perp b_\mathrm{rms})$ is built on the large-scale magnetic fluctuations, $t_{nl}=1/(k_\perp \sqrt{k_\perp \mathrm{E}_T})$ is calculated using the total energy spectrum, the exponent on the r.h.s indicates a fast ($\alpha=1$) or slow ($\alpha=2$) Alfv\'enic coupling, and $A$ is a constant of order 1.
The above spectral slopes at the peak of the turbulent activity suggest a different scenario for the HPIC ($\alpha=1$) and for the HMHD ($\alpha=2$) runs. 
By comparing the l.h.s. and the r.h.s. of Eq.~(\ref{eq:dyn}) at different times, we found that in HPIC the exponent $\alpha=1$ reproduces the slope of the ratio ${\mathrm{E}_R}/{\mathrm{E}_T}$ at all times past the peak of turbulent activity. On the contrary, in  HMHD, the exponent increases steadily from 2 to 3. 
This is due to the behavior of the parallel components in HMHD. In fact,
the exponent matching the slope of the ratio ${\mathrm{E}_R}/{\mathrm{E}_T}$ is stable also in HMHD when only the perpendicular components of magnetic and velocity fluctuations are used, and in the following we will restrict our analysis to those components (i.e., we redefine $\mathrm{E}_R = \mathrm{P}_{\boldsymbol{B}_\perp} - \mathrm{P}_{\boldsymbol{v}_\perp}$  and $\mathrm{E}_T = \mathrm{P}_{\boldsymbol{B}_\perp} + \mathrm{P}_{\boldsymbol{v}_\perp}$).
In Fig.~\ref{fig:dyn}, the l.h.s. of Eq.~(\ref{eq:dyn}), averaged for about 1.5 large-eddy turnover times, is shown for the HPIC (top) and the HMHD (bottom) in thick solid lines, along with the average of the r.h.s. with $\alpha=1,2$ (thin solid and dashed lines, respectively). 
The thick and thin solid lines are parallel in the MHD range ($0.3\lesssim k_\perp d_i\lesssim1$) for both simulations, thus supporting a fast scenario ($\alpha=1$), while the slow scenario ($\alpha=2$, dashed lines) has a steeper power law.
Note that although the ratio $\mathrm{E}_R/\mathrm{E}_T$ has a different power-law index (the ratio is compensated with different indexes in the two figures), the coefficients $A\approx3$ and $\alpha=1$ are the same in both simulations.

\begin{figure}[htb] 
\centerline{\includegraphics[width=\columnwidth]{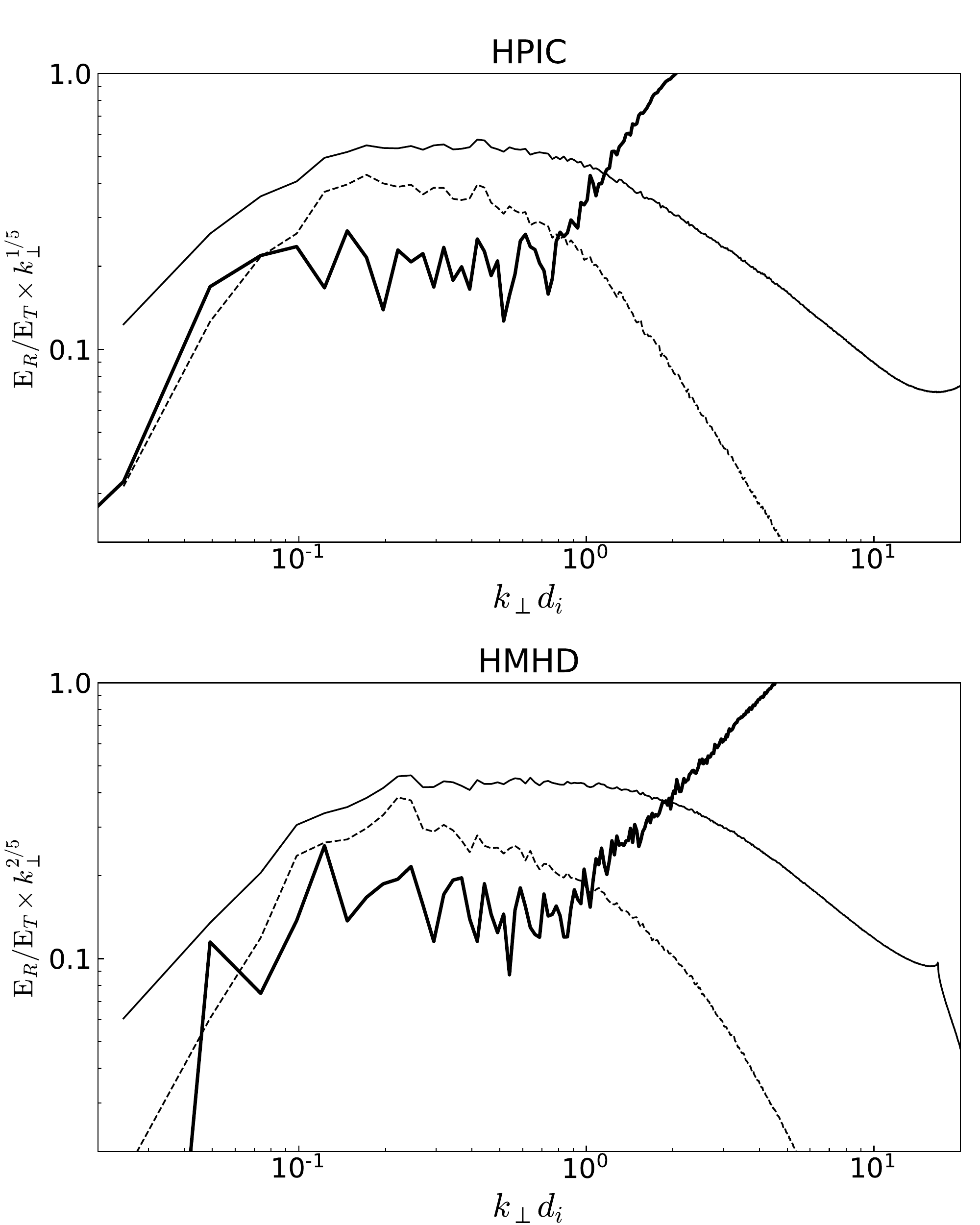}}
\caption{Ratio of the residual energy to the total energy (thick solid lines) in the perpendicular (2D) components for the HPIC (top panel) and HMHD (bottom panel) simulations, averaged over 60 times after the peak of current density. The ratio of the characteristic Alfv\'en time and dynamo time, $(t_A/t_{nl})^\alpha$ is plotted in thin solid and dashed lines for the fast and slow scenario, respectively ($\alpha=1$ and $\alpha=2$, see Eq.~(\ref{eq:dyn})).
 \label{fig:dyn}}
\end{figure}

Finally, we focus on the intermittent properties of the two simulations.
Localized current sheets and other coherent structures are intimately related to intermittency, that is the departure from Kolmogorov self-similarity law \citep{1941kolmogorov}. Among other properties, intermittency manifests itself with a non-gaussian behavior in the probability distribution functions (PDFs) of the increments of any field at a given scale $\ell$.
In Fig. \ref{fig:kurtosis}(a-c) we report the PDFs of the increments $\Delta B_y(x,y) = B_y(x+\ell,y) - \Delta B_y(x,y)$ of the $y$ component of the magnetic field fluctuations along the $x$ direction and at three different spatial separations $\ell/d_i=5\pi,\pi,$ and $\pi/4$, i.e., in the inertial range (a), at the spectral break (b), and at kinetic scales (c) respectively.
The PDFs are calculated at the same time $t=200~\Omega_i^{-1}$ in both simulations.
Indeed, the PDFs from the two models are comparable to each other at all scales. They are consistent with a Gaussian function at large scales in the inertial range ($\ell \gg d_i$), but as scales get smaller (e.g., already at $\ell =\pi d_i$ in Fig. \ref{fig:kurtosis}(b)), they depart from gaussian behavior and develop fatter and fatter tails.
\begin{figure}
    \centering \includegraphics[width=.86\columnwidth]{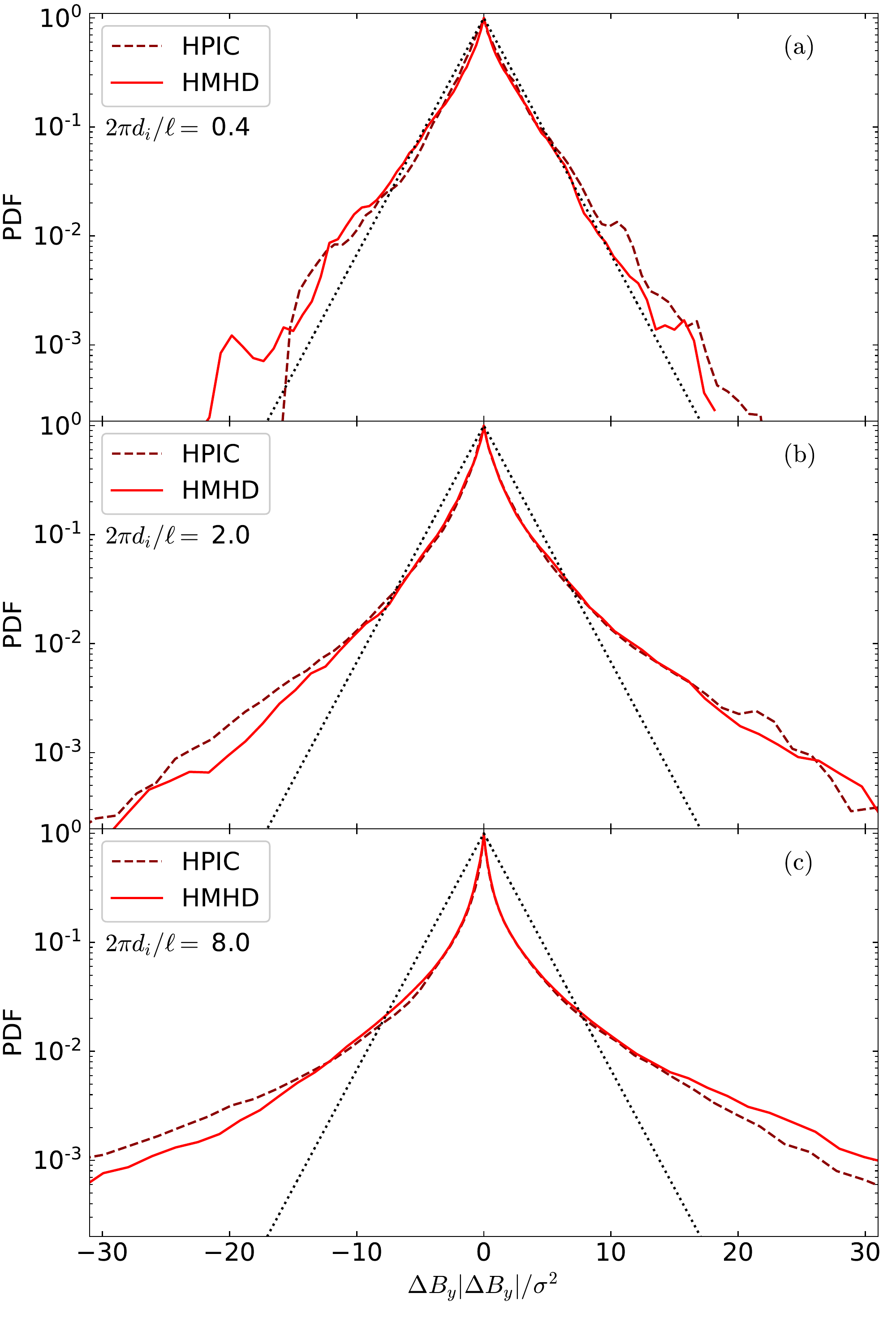}\\
    \includegraphics[width=.86\columnwidth]{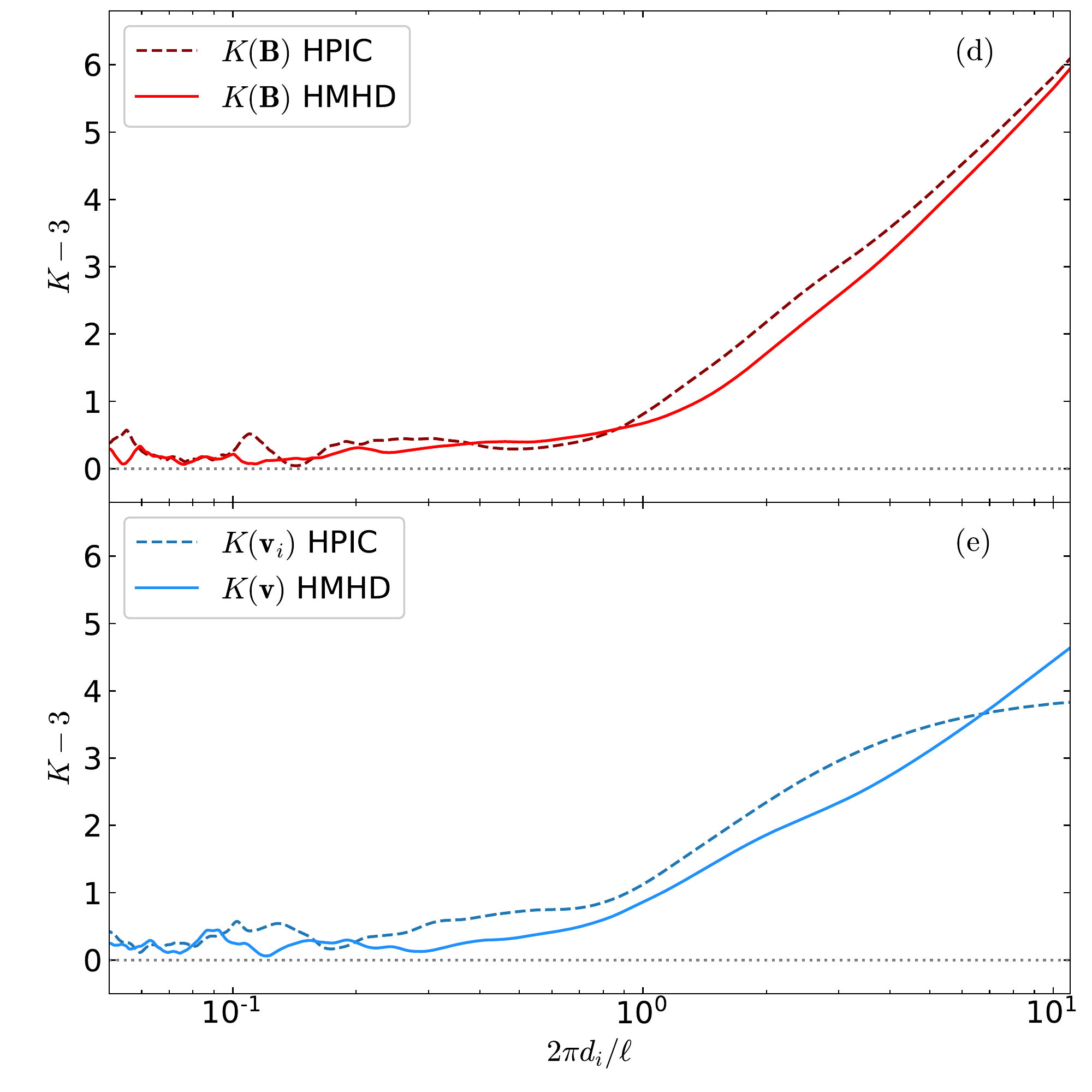}
    \caption{Top panels: PDFs of the increments $\Delta B_y = B_y(x+\ell,y) - \Delta B_y(x,y)$ at three different scales $\ell$: in the inertial range (a), at the spectral break (b), and at sub-ion scales (c). The dotted lines drawing a triangle correspond to a Gaussian function.
    The PDFs are plotted against
    $\Delta B_y |\Delta B_y|/\sigma^2$, with $\sigma^2$ being the variance of each PDF.
    Bottom panels: excess kurtosis $K - 3$ for the magnetic field (d) and the velocity field (e), at $t=200~\Omega_i^{-1}$ for both the HPIC run (dashed lines) and for the HMHD run (solid lines). The horizontal dotted line denotes the zero excess kurtosis of a Gaussian function.}
    \label{fig:kurtosis}
\end{figure}
To better quantify the level of intermittency, we calculated the scale dependent excess kurtosis, shown in Fig. \ref{fig:kurtosis} for the increments of the magnetic field (d) and of the velocity fields (e). The results for the magnetic field kurtosis are in remarkable agreement, being close to zero at large scales down to $2\pi d_i/\ell\approx 0.8$ and then increasing linearly down to the smallest scales where HPIC has a slightly larger kurtosis. 
The excess kurtosis of the velocity field increments are also similar at large scales,
but the HPIC kurtosis becomes slightly larger already at intermediate scales, $2\pi d_i/\ell \gtrsim 0.3$, well before the velocity spectra start to diverge ($k_\perp d_i \gtrsim 2$, see Fig. 4 (b)).

\section{Discussion}

In this work, we provided numerical evidence that many of the statistical and spectral properties of plasma turbulence at sub-ion scales can be explained within the framework of Hall Magnetohydrodynamics. 
The results have been obtained by performing a study of turbulence generated by freely decaying Alfv\'enic\CH{-like fluctuations
with zero mean cross helicity} using both a full viscoresistive fluid Hall-MHD model and a Vlasov-Maxwell hybrid particle-in-cell model coupling fully kinetic ions to fluid and massless electrons, which was able to correctly reproduce in situ observations \citep{2017franci_astronum}.
In the plasma regime we investigated, the Hall-MHD and the hybrid model showed a remarkable agreement.

In the early evolution of our simulations and long before the formation of a power-law direct cascade at fluid
scales, \CH{we observed, concurrently to the trigger of the first reconnection events, a transfer of energy from the large-scale vortices directly to small kinetic scales. After that, in a short transition phase, 
energy is fed  into the whole kinetic range, which we interpret as the signature in Fourier space of the coalescence of magnetic islands and their expulsion from the first current sheets into the turbulent surroundings. At later times, after few eddy turnover times, once turbulence is almost developed, we are not able to separate the contribution of magnetic reconnection to the power spectrum. However,} it is likely that such a reconnection-mediated energy transfer occurs whenever a current sheet reconnects, as suggested by \cite{2017franci} and corroborated by recent studies that employ advanced techniques to measure the scale-to-scale energy transfer in localized coherent structures \citep[e.g.][]{2017yang,2018camporeale}.
Our findings further strengthen the view that magnetic reconnection provides a direct channel to drive turbulence at kinetic scales \citep{2017franci}, and thus supports theoretical models of reconnection-mediated turbulence \citep{2017mallet,2017loureiro}.

\begin{figure}
 \includegraphics[width=\columnwidth]{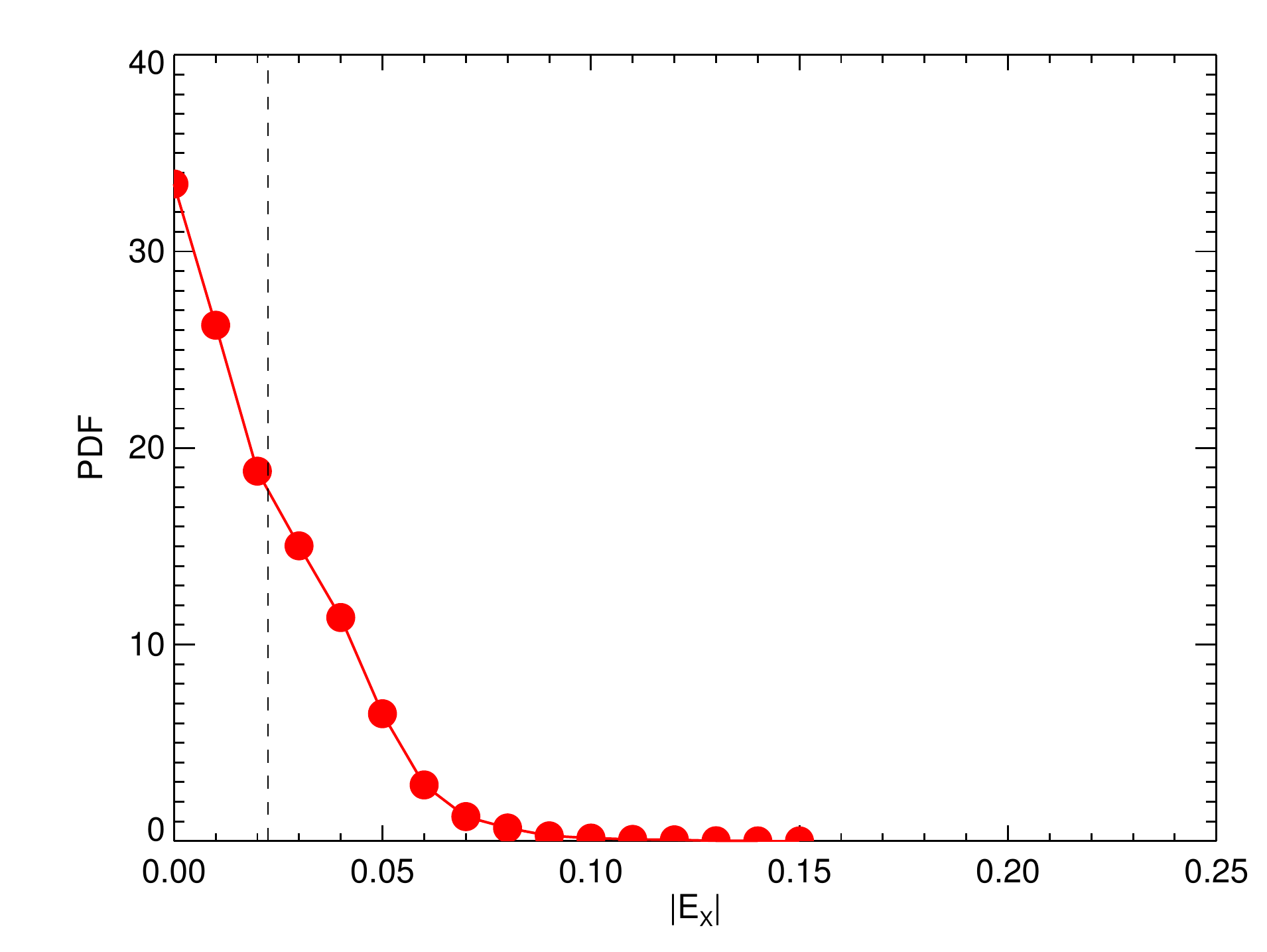}
 \caption{\CH{PDF  of $|\mathrm{E_X}|$ at the X-points, normalized with respect to the global Alfv\'en time $\tau_A=L_0/c_A$, with $L_0=256 d_i/2\pi$. Solid red lines and filled circles denote a linear binning (constant $\Delta|\mathrm{E}_X| = 0.01$) of the PDF. The vertical dashed line denotes the mean reconnection rate $\sim0.023\tau_A^{-1}$ of the distribution.}}
 \label{fig:ex_servidio}
\end{figure}

\CH{It is useful to compare our results on the reconnection rates with similar studies encompassing MHD \citep[][hereafter SV10]{2010servidio} and fully-kinetic PIC \citep[][hereafter HG17]{2017haggerty} simulations of turbulence.
In particular, our findings that the reconnection rates follow a lognormal distribution seems to contrast with the results of SV10 and HG17, who observed different distributions (see Fig. 7 of SV10 and Fig. 11 of HG17, respectively). To make a meaningful comparison with their work, we measured the values of the reconnecting out-of-plane electric field $\mathrm{E_X}=\partial_t A_z$ at the X-points (that is the definition of reconnection rate used by both SV10 and HG17) and renormalized the values to the global Alfv\'en time, defined on the global scale $L_0 = 256 d_i /2\pi$, as done in the above works.
Indeed, the PDF of $|\mathrm{E_X}|$ drawn in Fig. \ref{fig:ex_servidio} by using a linear binning with roughly the same binsize of SV10 is both qualitatively and quantitatively similar to the PDFs of SV10 and HG17.}

Once turbulence has fully developed, both the Hall-MHD and the hybrid model reproduce a Kolmogorov-like cascade of spectral index $-5/3$ at fluid scales and a kinetic cascade of spectral index $-3$ in the total magnetic field spectra. The location of the spectral break is also recovered. A flatter power-law spectrum of slope $-3/2$ is observed in the velocity spectra at fluid scales, 
which is also responsible for the different slope in the total energy spectrum ($-5/3$ and $-3/2$ in HPIC and HMHD, respectively). 
We also obtain a residual energy spectrum of slope $\approx-2$. These indexes are not consistent with the same Alfv\'en-dynamo balance regulating the ratio of residual and total energy at inertial range scales \citep{2005mueller,2016grappin}. However, when only perpendicular component are retained (i.e. 2D wavevectors and components are analysed), both HPIC and HMHD display properties consitent with a fast Alfv\'en-dynamo balance scenario ($\alpha=1$).

HPIC and HMHD simulations show a different behaviour of the velocity field spectrum at ion-scales the latter being flatter, and with a stronger Alfv\'enic coupling in the transition from MHD to ion scales.
The overall agreement extends to other statistical properties, related to intermittency, and further confirms that the Hall-MHD model may be accurate enough to describe the dynamics of magnetic fields fluctuations in a turbulent plasma.

Results of this work suggest that the Hall term is the dominant term shaping the turbulent properties and dynamics of the magnetic fields at sub-ion scales.
That rules out other mechanisms not described by our Hall-MHD model, for instance gyrotropic effects, temperature anisotropies, particle-wave interaction effects such as Landau damping, cyclotron resonance, and linear Vlasov instabilities.
Some of these effects are likely responsible for the differences arising between the HMHD and the HPIC model in the power spectra of the velocity fluctuations at kinetic scales. In particular, at large scales, the difference between the HMHD and HPIC  power spectra of parallel magnetic field and velocity fluctuations may be due to nonzero off-diagonal terms in the ion pressure tensor generated by non-gyrotropic effects at the reconnection sites \citep{2002yin}, which may also be responsible for the differences between the reconnection rates of the Hall and the hybrid model. 

\CH{Recently, Gonzalez et al. investigated turbulence at kinetic scales by means of both a two-fluid incompressible Hall-MHD model retaining electron inertia effects and a fully kinetic plasma model. Although the use of a proton-to-electron mass ratio of 25 in those models hinders a direct comparison with our results in the sub-ion range, they also find an agreement in spectral and turbulent properties between the two models, even at electron scales. Their results further confirm the potential use of Hall-MHD fluid models in the study of many properties of turbulence at kinetic scales.
}

The ability of Hall-MHD of describing the plasma dynamics at sub-ion scales needs to be further investigated in other plasma regimes.
A more detailed analysis highlighting the differences between HMHD and HPIC simulations may provide further constraints on additional phenomena
beyond the Hall physics relevant for plasma turbulence at sub-ion scales.
\CH{For instance, preliminary results from analyses based on the von Karman-Howarth equations 
show interesting similarities, but also differences, between HMHD and HPIC simulations, thus complementing the present work. This will be the subject of future works.}

\begin{acknowledgments}
 
The authors wish to thank William H. Matthaeus for useful discussion.
This research was partially supported by the UK Science and Technology Facilities Council (STFC) grant ST/P000622/.
This work was supported by the Programme National PNST of CNRS/INSU co-funded by CNES.
We acknowledge partial funding by Fondazione Cassa di Risparmio di Firenze under the project HIPERCRHEL.
P.H. acknowledges grant 18-08861S of the Czech Science Foundation.
The authors acknowledge the ``Accordo Quadro INAF-CINECA (2017)'', for the availability of high performance computing resources and support, PRACE for awarding access to the resource Cartesius based in the Netherlands at SURFsara through the DECI-13 (Distributed European Computing Initiative) call (project HybTurb3D), and CINECA for awarding access to HPC resources under the ISCRA initiative (grants HP10B2DRR4 and HP10C2EARF). 
\end{acknowledgments}

\appendix
\section{Reconnection rates: Identification of magnetic X/O-points.}
\CH{
Characterizing the statistical properties of reconnection at a given time, requires the measurements of several physical quantities at the magnetic X- and O-points inside all current sheets present in the simulation. 
In 2D models, these points are the critical points (i.e., minima, maxima, and saddles) of the out-of-plane vector potential $A_z$.
To that purpose, we designed an algorithm that, for each output of the simulation, performs the following operations:
\begin{enumerate}
 \item Identification of critical X/O-points of $A_z$;
 \item Calculation of $A_z$, $\partial_t A_z$, and other ancillary quantities at the critical points;
 \item Identification of current sheet structures, if present;
 \item Selection and pairing of the X/O-points located inside each current sheet;
 \item Calculation of the reconnection rate of each XO-pair.
\end{enumerate}
The algorithm is implemented in IDL, and employs Fourier
decomposition to calculate all derivatives.  
The first two steps are common to other algorithms used for the analysis of magnetic reconnection in turbulence \citep[e.g.,][]{2010servidio,2017haggerty}. In those studies, the reconnection rate is defined as the value of $\partial_t A_z/b_\text{rms}^2$ at the X-points,  normalized with respect to the mean-square magnetic field $b_\text{rms}^2$. %Moreover, regardless of whether they are located in current sheet structures. 
Instead, we use the following definition (see Eq. (\ref{eq:recrate}))
\begin{equation}
\label{eq:recrate_appendix}
 \gamma_\mathrm{rec} = \left |\frac{1}{\Phi|_X^O} \frac{\partial \Phi|_X^O}{\partial t} \right |= \left |
 \frac{1}{A_z^O - A_z^X} \partial_t(A_z^O - A_z^X)\right |,
\end{equation}
which self-consistently defines a quantity with the desired dimensions (i.e., the inverse of a time), without the need for any further normalization. This definition is also consistent with the classic definition of the growth rate adopted for the tearing instability, as shown in a previous study \citep{2018papini}.
The above steps are implemented as follows.}

\textbf{Step 1}: 
\CH{
We begin by taking the components of the gradient of $A_z$, which in our models correspond to the in-plane components of the magnetic field
\begin{equation}
 \partial_x A_z = - B_y \,,\quad \partial_y A_z = B_x \,.
\end{equation}
We then compute the zero contour lines of $B_x$ and $B_y$, by using the IDL \texttt{ISOCONTOUR} procedure. The critical points are the intersection points of such lines, where $|\nabla A_z |= 0$. 
We found that this method is less prone to the detection of spurious critical points \citep{2010wan} than the use of other interpolation techniques.
The remaining operations are analogous to \citet{2010servidio}. For each of the critical points, we compute the determinant and the eigenvalues of the Hessian matrix 
\begin{equation}
 \mathcal{H}(x_i,y_i) =\left |\begin{array}{ll}
  \partial_x^2 A_z(x_i,y_i) & \partial_x\partial_y A_z(x_i,y_i) \\
  \partial_x\partial_y A_z(x_i,y_i) & \partial_y^2 A_z(x_i,y_i)
 \end{array}\right |
\end{equation}
at the coordinates $(x_i,y_i)$ of the critical point (which are usually located between the grid points), by means of bilinear interpolation of the derivatives of $A_z$ at $(x_i,y_i)$.
Finally, positive and negative values of $\det(\mathcal{H})$ identify O-points and X-points respectively.  
}

\textbf{Step 2}: 
\CH{$A_z$ is given as output by the HMHD code, while in CAMELIA is calculated via Fourier integration of the magnetic fields $B_x$ and $B_y$. The time derivative, $\partial_t A_z$, is calculated by taking the $z$ component of the induction equation for the vector potential
\begin{equation}
\partial_t A_z = (\boldsymbol{v} \times \boldsymbol{B})_z - \eta (\nabla \times \boldsymbol{B})_z  -\eta_H \frac{((\nabla \times \boldsymbol{B}) \times \boldsymbol{B})_z}{\rho}.
\end{equation}
Finally, the values at the X/O-point are found via interpolation. 
}

\textbf{Step 3}: 
\CH{
Current sheets in the simulations are defined by the regions where the amplitude of the current density is larger than a given threshold $\epsilon_0$. The choice of $\epsilon_0$ is somewhat arbitrary. For our purposes, a good choice is to set $\epsilon_0$ equal to the maximum between $10\%$ of the maximum amplitude of $\boldsymbol{J}$ in the simulation box and ten times its mean, at each different time. 
}

\textbf{Step 4}: \CH{We select the X-points belonging to each current sheet separately and we pair them with the nearest O-point belonging to the same current sheet. We discard the pairs whose distance is less than four grid points.}

\textbf{Step 5}: \CH{Using Eq. (\ref{eq:recrate_appendix}), we obtain the reconnection rate of each XO-pair from the values of $A_z$ and $\partial_t A_z$ calculated in step 2.}

 \bibliographystyle{apj2-eid}
 \bibliography{bibliography}
\end{document}